\documentclass[letterpaper]{article}

\usepackage[T1]{fontenc}

\usepackage{geometry}
\usepackage{setspace}

\usepackage[
  backend=biber,
  style=chem-acs,      
  sorting=none,
  natbib=true,
  sortcites=true,
  articletitle=true,
  doi=true
]{biblatex}
\renewbibmacro{in:}{}

\usepackage[
  colorlinks=true,
  linkcolor=blue,
  citecolor=blue,
  urlcolor=black
]{hyperref}

\usepackage{graphicx}
\usepackage{float}
\newfloat{scheme}{htbp}{los}
\floatname{scheme}{Scheme}
\floatname{chart}{Chart}
\newfloat{graph}{htbp}{loh}

\usepackage{chemformula} 
\usepackage[version = 4]{mhchem} 

\usepackage{tabularx,makecell}
\usepackage{amsmath, amssymb, bm, graphicx, caption, subcaption}
\usepackage{geometry}
\usepackage{physics}
\usepackage{siunitx}
\DeclareSIUnit\angstrom{\text{Å}}
\usepackage{float}
\usepackage{tikz}
\usepackage{dblfloatfix}
\usepackage{lipsum}
\usepackage{authblk}
\usepackage{caption}
\usepackage[colorlinks=true,linkcolor=blue, citecolor=blue]{hyperref}
\usepackage{xcolor}

\usepackage{xr}
\usepackage{hyperref}

\usepackage{authblk}

\author[1]{Pablo Res\'endiz-V\'azquez\thanks{Corresponding author: pab.rv94@gmail.com}}
\author[2,3]{Christophe de Beule\thanks{Corresponding author: christophe.debeule@uantwerpen.be}}
\author[1]{Thi-Hai-Yen Vu}
\author[1]{Kaijian Xing}
\author[1]{Daniel McEwen}
\author[4,5]{Daniel Bennett}
\author[6]{Liangtao Peng}
\author[7,8]{H\'ector Gonz\'alez-Herrero}
\author[6]{Shaffique Adam}
\author[1]{Mark T. Edmonds}
\author[1]{Michael S. Fuhrer\thanks{Corresponding author: michael.fuhrer@monash.edu}}

\affil[1]{\small School of Physics and Astronomy, Monash University, Clayton, Victoria 3800, Australia}
\affil[2]{\small Department of Physics and Astronomy, University of Pennsylvania, Philadelphia, Pennsylvania 19104, USA}
\affil[3]{\small Department of Physics, University of Antwerp, Groenenborgerlaan 171, 2020 Antwerp, Belgium}
\affil[4]{\small John A.~Paulson School of Engineering and Applied Sciences, Harvard University, Cambridge, Massachusetts 02138, USA}
\affil[5]{\small School of Electrical and Electronic Engineering, Nanyang Technological University Singapore, 50 Nanyang Avenue, 639798, Singapore}
\affil[6]{\small Department of Physics, Washington University, St. Louis, Missouri 63130, USA}
\affil[7]{\small Departamento de F\'isica de la Materia Condensada, Universidad Aut\'onoma de Madrid, E-28049 Madrid, Spain}
\affil[8]{\small Condensed Matter Physics Center (IFIMAC), Universidad Aut\'onoma de Madrid, E-28049, Madrid, Spain}

\graphicspath{{./Figs/}}

\title{Disorder-induced symmetry breaking in moir\'e bands of marginally twisted bilayer MoS$_2$}

\date{}

\begin{document}

\maketitle

\begin{abstract}
  Twisted transition‑metal dichalcogenides host highly tunable 
  moir\'e potentials, flat bands, and correlated electronic
  phases, yet the role of disorder in shaping these emergent
  properties remains largely unresolved. 
  Using scanning tunneling spectroscopy, we investigate the 
  impact of electrostatic disorder on the electronic structure of 
  marginally twisted ($\theta \approx 0.95^\circ$) bilayer 
  MoS$_2$. 
  Differences of 15 meV in the onset energies of the valence and
  conduction bands between MX and XM stacked regions are 
  observed, and are unexpected due to symmetry considerations. 
  We also observe spatially correlated disorder in the onset 
  energy that is consistent with a background random charge
  density of a few $10^{11}$ $\text{cm}^{-2}$. Continuum model calculations for twisted MoS$_2$ show dramatic changes in the low-energy moir\'e bands in response to an electric displacement field, in quantitative agreement with experiment. Moreover, the calculated local density of states with disorder smearing compares well to experiment only when structural relaxation is accounted for.
  These results highlight the critical role of electrostatic
  disorder in the electronic structure of moir\'e materials at 
  the nanoscale.
\end{abstract}




In twisted superstructures (moiré superlattices), such as 
twisted transition metal dichalcogenides (tTMDs), electron 
interference gives rise to nearly non-dispersing ``flat'' bands. 
These flat bands are known to host strongly correlated 
electronic phenomena, including generalized Kane–Mele 
phases~\cite{kane_mele_exp}, fractional quantum anomalous Hall 
states~\cite{crommie_qah}, generalized Wigner 
crystals~\cite{crommie_wigner}, Haldane-
type~\cite{zhao2024realization} and Mott-type insulating 
phases~\cite{guo2025mott}, superconducting 
states~\cite{guo2025superconductivity} and topologically non-
trivial phase transitions~\cite{topology_tmds}.

Unlike the paradigmatic case of twisted bilayer graphene, where 
a single \textit{magic} angle is required for flat bands and the 
observation of correlated electron states~\cite{jarillo1, 
jarillo2}, in small-angle tTMDs, theoretical studies predict 
that the bandwidth of the low-energy moir\'e bands decreases 
continuously with twist angles near both parallel ($0^\circ$) and 
antiparallel ($60^\circ$) stacking~\cite{small_angle_var, 
tunning_bands}, making them an accessible platform to 
investigate a rich variety of emergent quantum 
phases~\cite{marvels_moire, review_tmds}.

Although crystallographic growing techniques have considerably 
advanced in producing high-quality crystals~\cite{liu2023two}, 
intrinsic disorder persists as an unavoidable feature in two-
dimensional (2D) materials~\cite{holbrook2025revealing, 
xu2023validating}. 
This disorder has been shown to have a strong influence on the 
electronic properties of these materials, such as modifying 
the interlayer coupling through local charge 
redistribution~\cite{disorder_moire}, altering lattice 
relaxation via inhomogeneous local strain 
fields~\cite{local_strain}, limiting the mobility of charge 
carriers~\cite{zhang2009origin} or causing Fermi energy 
fluctuations in Dirac semimetals~\cite{edmonds2017spatial} due 
to the formation of charge puddles, and even driving quantum 
phase transitions~\cite{holbrook2024real, crommie_dis, 
defect_assisted}.

In TMDs, the formation mechanisms and microscopic nature of 
disorder, as well as its role in the electronic structure, 
remains under debate~\cite{mos2_intr_dis_stm, 
holbrook2025revealing}. 
This naturally raises the question of whether disorder could 
also play an important role in tTMD moir\'e systems.

While angle-resolved photoemission spectroscopy 
(ARPES)~\cite{arpes1,feraco2024nano} and electrical transport 
measurements~\cite{transport1, transport2} have been useful in 
measuring the band structure of these materials, unambiguously associating disorder with local electronic structure at the nanoscale requires a local probe such as scanning tunneling microscopy (STM) and spectroscopy (STS)~\cite{wse2_twisted,eva_andrei_pol,special_angle}. 
These techniques offer a unique way to measure the local density 
of states (LDOS) that populate such bands as well as its spatial 
modulation across distinct moir\'e regions~\cite{marvels_moire}.

STM/STS studies have revealed key signatures in the LDOS of 
small-angle tTMDs. 
Notable recent examples include the observation of flat bands at 
the $\Gamma_{v}$ point of the first moiré Brillouin zone (mBZ) 
of antiparallel twisted bilayer WSe$_2$ over a range of marginal 
twist angles~\cite{wse2_twisted}; 
the measurement of an LDOS shit close to the 
valence band maxima between highly symmetric regions in parallel 
twisted bilayer MoS$_2$~\cite{eva_andrei_pol}, attributed to the 
inverse intrinsic electric polarization of these stackings; 
and the identification of a ``special angle'' in twisted bilayer 
MoS$_2$ at $1.7^\circ$, where the bandwidth of the ``flat'' band 
at $\Gamma_{v}$ reaches a local minimum value of 
$195$~meV~\cite{special_angle}. 
However the role of native disorder in the moir\'e bands has 
received little attention.

In this work, we investigate the impact of electrostatic 
disorder on the moir\'e band structure of marginally twisted 
($\theta \sim 0.9^\circ$) bilayer MoS$_2$ (tb-MoS$_2$) using 
STM/STS. 
Topographic measurements reveal a reconstructed triangular 
moir\'e superlattice with pronounced spatial disorder. 
Local spectroscopic measurements across distinct stacking 
regions show an unexpected splitting of the onset energy of the 
valence and conduction band edges between mirror symmetric 
moir\'e sites. 
We attribute these observations to an interlayer electrostatic 
potential induced by randomly distributed charged defects, 
likely sulfur vacancies. 
By developing an electrostatic disorder model, we infer a 
background charge density consistent with sulfur vacancy 
densities reported for exfoliated monolayer 
MoS$_2$~\cite{mos2_intr_dis_stm, lu2014bandgap, 
mcdonnell2014defect,exp_vac}. 
We then perform continuum-model calculations of the moir\'e band 
structure under symmetry-breaking out-of-plane electric fields 
consistent with the inferred disorder potential. 
These calculations reveal a pronounced splitting of the low-
energy moir\'e bands near the valence and conduction band edge, 
in excellent agreement with our experimental results.

\section{Results}

\begin{figure*}[ht!]
    \centering
    \includegraphics[width=1.0\textwidth]{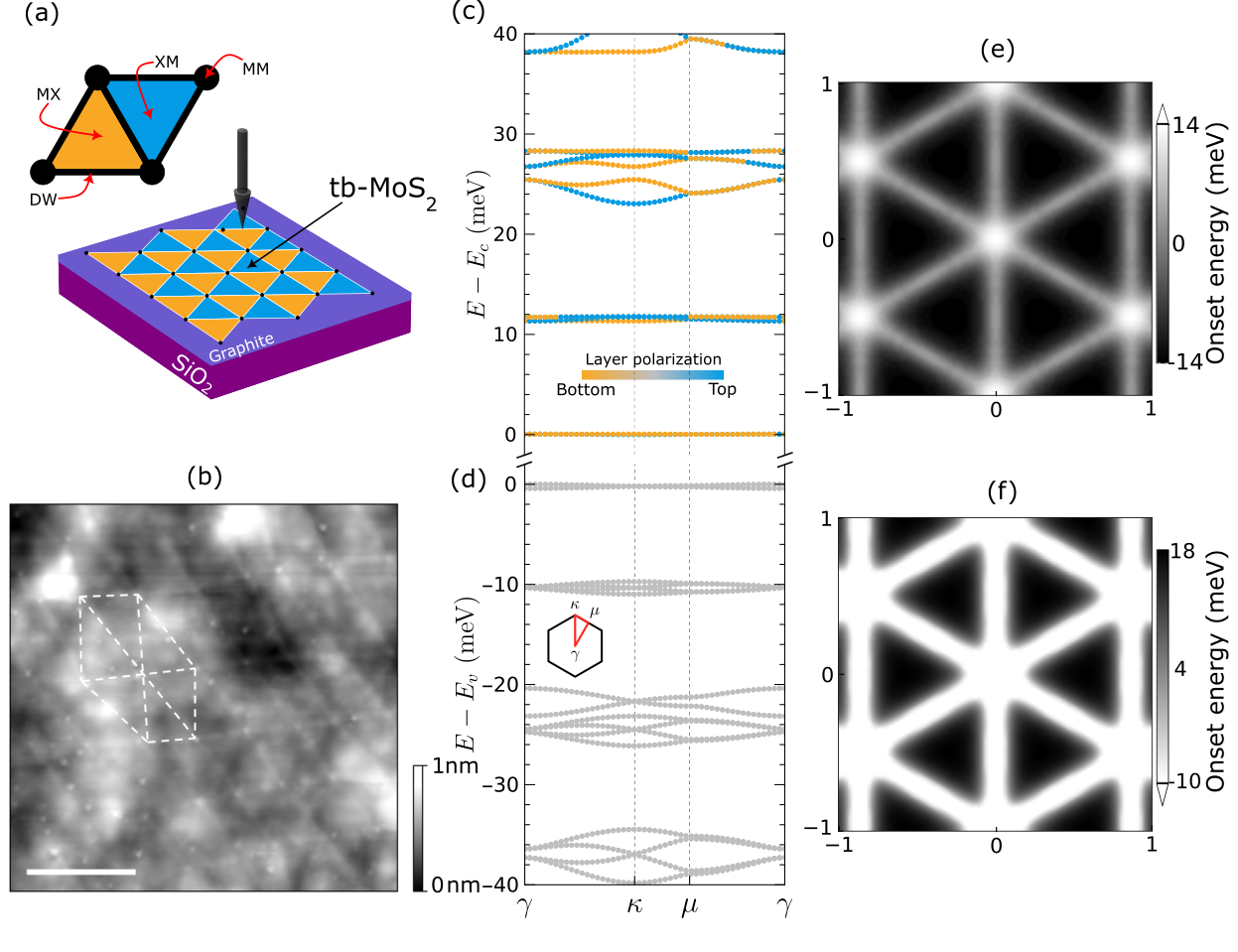}
    \caption{
    (a) Schematic of the tb-MoS$_2$ device on a graphite/SiO$_2$ substrate. 
    The inset shows the structure in a moir\'e unit cell, with MM sites, MX/XM stacking domains, and DWs. 
    (b) Large area STM topograph of the tb-MoS$_2$ device showing a moir\'e lattice with a lattice constant of $\sim18$~nm corresponding to a twist angle of $\sim 0.95^\circ$. A dotted white hexagon outlines the domain walls (DW) of a few moir\'e unit cells. 
    Scale bar: $40$~nm. 
    Measurement parameters: $V_{\text{bias}}=-1.9$ V, $I_{\text{set}}=100$ pA. 
    (c,d) Low-energy bands computed with the continuum model for (c)the conduction and (d) the valence band, along the path $\gamma\rightarrow\kappa\rightarrow\mu$, in the mBZ shown in the inset.
    The projection of the layer polarization onto the bands is shown, as indicated by the color scale.
    (e,f) Calculated spatial maps of onset energies for conduction (e) and valence (f) bands showing a honeycomb hexagonal lattice similar to one observed in (a). 
    Given that for some locations the energy intersection with the arbitrarily-set LDOS value does not exist within the energy range, the color bars are unbounded in one direction (indicated by the arrow).}
    \label{fig:fig1}
\end{figure*}

Figure.~\ref{fig:fig1} provides an overview of the experiment.
Figure~\ref{fig:fig1}a schematically illustrates the 
experimental device setup. 
Samples were fabricated using the tear-and-stack technique (see 
Methods section and Supporting information (SI) Figure~S1 for more details). 
Using the STM probe, we observe a large moir\'e superlattice 
that emerges when two marginally twisted MoS$_2$ monolayers 
undergo lattice relaxation. 
This moir\'e structure consists of a hexagonal honeycomb lattice 
formed by two oppositely arranged triangular domains made of MX 
(orange) and XM (blue) stacking registries.
Here, M (X) represents the Mo (S) atoms, with the letter order 
indicating the vertical stacking sequence. 
Thin domain walls (DW) separate the MX/XM regions, while MM 
registries, for which the Mo atoms of different layers are 
directly on top of each other, are confined to the vertices of 
the 
lattice~\cite{carr_relaxation_2018,relaxation_theo_1,debeule_theory_2025}, as illustrated in the inset. 

Figure~\ref{fig:fig1}b shows an STM topograph of the tb-MoS$_2$ 
moir\'e superlattice. 
The MM regions appear as bright dots, and the domain walls as 
bright lines. The observed average periodicity is $\lambda_{M} 
\simeq 18.7 \pm 0.5$ nm, using $ \text{a} = 0.31$~nm for the 
MoS$_2$ lattice constant~\cite{mos2_lattice}, 
yielding a twist angle of $\theta \sim 0.95^\circ \pm 
0.03^\circ$.

The scan also reveals topographic disorder that 
appears correlated on length scales exceeding the moir\'e period, as well as 
heterostrain (distortion of the moir\'e lattice) inherent to the morphology of the sample region. 
To highlight the superlattice we overlay a dotted white hexagon 
as a visual guide. 

The band structure of tb-MoS$_2$ was calculated using a 
continuum model (see Methods and SI for 
details), with the conduction and valence bands shown in 
Figures~\ref{fig:fig1}c and d, respectively, for $\theta \sim 
0.95^\circ$.
The effects of atomic relaxation 
\cite{nam_lattice_2017,carr_relaxation_2018,bennett_theory_2022,debeule_theory_2025} were included in the continuum model (see 
Methods and SI for details).
The vertical energy scales are measured relative to the energies 
$E_{c}$ ($E_{v}$) of the lowest (highest) energy moir\'e 
miniband derived from the MoS$_2$ conduction (valence) band, 
while the color scale represents the layer polarization between 
the top and bottom MoS$_2$ layers. 
Narrow, few-meV-wide bands are observed near the band edges of 
both conduction and valance bands.

In the conduction band, the first moir\'e band manifold consists of two nearly degenerate subbands, with weights on the top and bottom layers, respectively. 
These bands predominantly originate from Mo-$d_{z^2}$ orbitals with only minor contributions from S-$p_{x}$ and S-$p_{y}$ orbitals~\cite{zhu2016evolution}. 
The interlayer coupling between the top and bottom layers is strong for the MM stacking, but relaxation shrinks these regions to points, hence weak interlayer tunneling results in near-degeneracy of the two subbands (see SI~Figure~S10b).

In contrast, the valence bands are strongly layer-hybridized. 
In this case the Wannier orbitals, which originate mostly from the Mo-$d_z^2$ and S-$p_z$ orbitals~\cite{special_angle,eva_andrei_pol, small_angle_var, zhu2016evolution}, are a superposition of states from both layers, coupled through strong interlayer tunelling at MX and XM regions, and the two topmost (spin degenerate) valence bands have a graphene-like character with bonding and anti-bonding subbands crossing at Dirac points at $\kappa/\kappa^{\prime}$ in the mBZ. (see SI~Figure~S10a). 

In order to visualize the spatial variation in LDOS corresponding to the moir\'e bandstructure, in Figures~\ref{fig:fig1}e and f we plot "onset maps", these are spatial distributions of the energies at which the LDOS first rises above an arbitrarily-set value for the conduction and valence band edges, respectively. 
Here, the LDOS is computed from the moir\'e continuum model for the top (bottom) of the valence (conduction) band where we sum over both layers of tb-MoS$_2$ (see Methods section for further details). 
These onset maps feature the same hexagonal moir\'e lattice reconstruction observed in the topographic 
measurements.
We can see that in the case of pristine tb-MoS$_2$, the LDOS shows that the MX and XM states have the same onset energies, which is enforced by a mirror symmetry.

\begin{figure*}[ht!]
    \centering
    \includegraphics[width=1.0\textwidth]{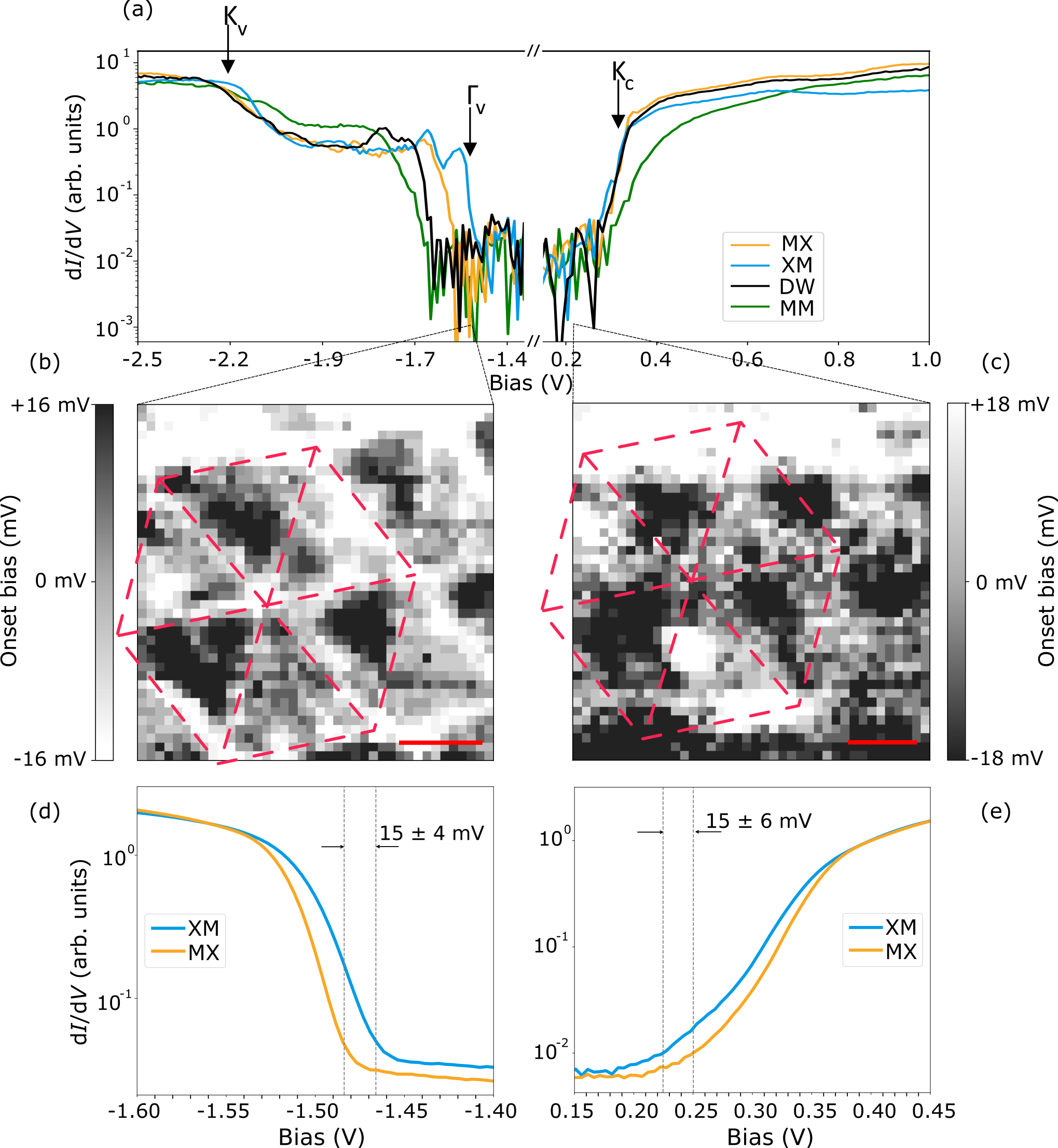}
    \caption{ (a) STS point measurements at the AA, MX, XM and DW regions of the  moir\'e lattice. Measurement parameters are $V_{\text{bias}} = -2.5 $~V , $I_{\text{set}} = 250 $~pA. Each curve is the average of 10 sweeps taken using lock-in amplifier with modulation amplitude of $10$~mV and frequency of $793$~Hz. (b,c) Onset maps over a $\sim 50 \times 50 $~nm$^2$ moir\'e region for the valence band (b) and conduction band (c) edges. Red dashed lines highlight some moir\'e domain walls. Scale bar is $10$~nm. (d,e) Spatially averaged LDOS for the MX/XM regions shown in (b) and (c). Vertical dotted lines show the calculated onset energies for each curve, the absolute difference between them and its uncertainty are indicated in each plot. The vertical axis of all the STS curves are in logarithmic scale.}
    \label{fig:fig2}
\end{figure*}

We now measure the electronic structure at the four high-symmetry stacking regions (MM, MX, XM and DW) of the moir\'e lattice with scanning tunneling spectroscopy (STS), in which the dI/dV spectrum (the differential conductance dI/dV as a function of sample bias V) is proportional to the LDOS at energy $E_{F} + e\text{V}$, where $E_{F}$ is the Fermi energy. Fig.~\ref{fig:fig2} shows our results. 

Figure~\ref{fig:fig2}a shows point spectra taken at the high-symmetry stacking regions of the moir\'e lattice. 
We observe three main LDOS features: (1) the onset of the valence band LDOS at roughly $-1.6$~V (labeled $\Gamma_{\text{v}}$), (2) the onset of the conduction band LDOS at roughly $+0.3$~V (labeled K$_{\text{c}}$), and (3) an additional onset of LDOS at approximately $-2.2$~V (labeled K$_{\text{v}}$). 
Through comparison to previous experimental and theoretical work~\cite{special_angle, twist_bands, theo_sheer, theo_elec_str}, we identify these LDOS features as corresponding to the valence band ($\Gamma_{\text{v}}$) and conduction band (K$_{\text{c}}$) edges which arise from states located near the $\Gamma$ and K/K' points of untwisted bilayer MoS$_{2}$, and the onset of LDOS associated with states located near K/K' in the valence band (K$_{\text{v}}$) . 
No dI/dV signal above the noise floor was observed in the bulk band gap and therefore in order to display the changes in the valence and conduction band edges, the measurements in the gap region were removed. 

The valence band edge at $\Gamma_{\text{v}}$ exhibits several notable features which are analyzed further below. 
The STS spectra acquired at MM and DW stacking regions show a pronounced shift towards lower bias compared to those measured in the MX and XM domains. 
On the other hand, the onset energy at MM is shifted towards lower bias relative to the DW onset, similar to the STS characteristics reported in Refs.~\cite{eva_andrei_pol, special_angle}.

The STS spectra at MX and XM stackings appear to show a shift in onset of LDOS at $\Gamma_{\text{v}}$, visibly more negative at MX than at XM. As discussed above, since the minibands are predicted to be completely unpolarized, such a shift is unexpected based on the symmetry of the pristine moir\'e system and it has not been reported in previous works. It is less clear from these point spectra whether a similar onset shift is observed at the conduction band edge at K$_{\text{c}}$.

In order to further explore and quantify the apparent spatial modulation of the LDOS onsets within several moir\'e unit cells, we performed STS maps over a region containing around 10 MX/XM domains (see Methods section for further measurement details). 
From these grid measurements, we extracted onset maps of the onset bias of the LDOS features $\Gamma_{\text{v}}$ and K$_{\text{c}}$. 
Similar to the theoretically obtained, these values are determined by finding the intersection between a noise floor and a linear fit to $\log{\left(dI/dV\right)}$ at the corresponding band edges, following the approach of Refs.~\cite{tang2017quantum,ugeda2016characterization,thomas2024substitutional} (see Methods section and SI~Figure~S2 for further details).
The results are shown in Figures~\ref{fig:fig2}b and c, respectively. 
In both onset maps, alternating dark and bright triangles with opposite orientation (highlighted by red dashed lines) can be distinguished. 
The spatial period matches the one of the moir\'e lattice obtained from the topography image in Figure~\ref{fig:fig1}b. 
Thus, the triangular shapes are identified as MX/XM domains. 
Substantial randomly localized onset bias variations at a scale similar to the color contrast between MX/XM regions is also seen in both maps, which we identify as disorder. 

To characterize the average onset shift between several moir\'e domains, spatially averaged STS curves of the same moir\'e regions (XM or MX) are obtained. 
The resulting spectra, shown in Figures.~\ref{fig:fig2}d and e, reveal a splitting of $15$ $\pm$ $4$ mV between MX and XM sites for the valence band and of $15$ $\pm$ $6$ mV for the conduction band (see Methods for further details).

\begin{figure*}[ht!]
    \centering
    \includegraphics[width=1.0\textwidth]{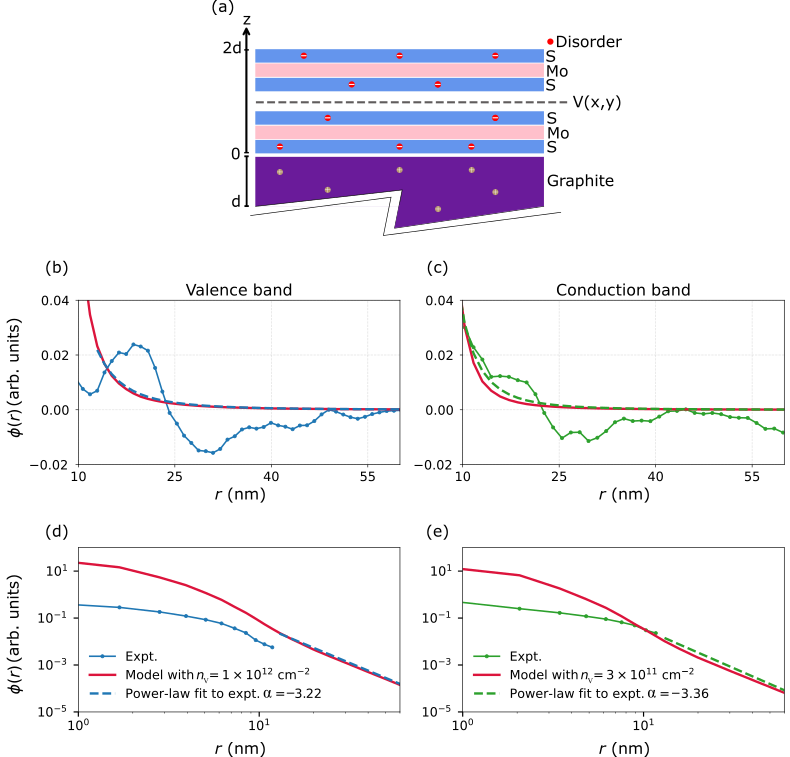}
    \caption{(a) Schematic of the electrostatic model used to estimate the electrostatic potential due to randomly-distributed charges in the sample; $d$ is the thickness of a MoS$_2$ monolayer and $V(x,y)$ is the total electrostatic potential. Red dots represent randomly distributed V$_{S}$ while brown dots inside the section of the graphite layer are the primary image charges. (b-e) Radially averaged disorder potential $\phi(r)$. Blue and green solid points represent the experimentally measured disorder potential $\phi(r)$, and dashed lines are power-law fits to the experimental data for $r > 10$ nm, for valence band (b,d) and conduction band (c,e) onset maps, respectively. Power-law exponents are given in the legends of d, e. Red lines in b-e are the electrostatic models that best fit the experimental profiles. In d and e we present the results for $r > 0$ nm in log-log scale with the obtained surface charge defect density, $n_{V}$ in the legends. }
    \label{fig:fig3}
\end{figure*}

In order to qualitatively analyze the role of disorder in our measurements, we developed an electrostatic model that allowed us to infer the density of trapped charges and the electric field associated with it. Experimental studies have reported the presence of point defects in exfoliated~\cite{mos2_intr_dis_stm} and in chemical vapor deposited (CVD) MoS$_2$ monolayers~\cite{zhou2013intrinsic}. Among other types of defects, sulfur vacancies (V$_{S}$) have been predicted by first‐principles calculations~\cite{native_def} and confirmed by STM and TEM experiments~\cite{mos2_intr_dis_stm, exp_vac} to be the most abundant lattice defects~\cite{mos2_intr_dis_stm, defects_vacancies}. These V$_{S}$ centers are stable in neutral or negatively charged states and therefore act as electron acceptors~\cite{native_def}. We assume that our tb-MoS$_{2}$ has an areal concentration $n_{\text{V}}$ of negatively-charged V$_{S}$ randomly distributed in the four S layers, as depicted in Figure~\ref{fig:fig3}a.

Figure~\ref{fig:fig3}a presents a schematic illustration of the electrostatic disorder model. To obtain the electrostatic potential due to the charges with concentration $n_{\text{V}}$, we treat the graphite substrate as an ideal metal. This is a reasonable approximation given that thin graphite has a Thomas–Fermi screening length of 5–\SI{7}{\angstrom}~\cite{graph_scre}, which is shorter than the thickness of the twisted‐MoS$_2$ bilayer ($\sim$\SI{12}{\angstrom}~\cite{bilayer_thick}).  We then use the textbook method of images~\cite{jackson_image}, applied to a system of alternating thin dielectric layers between a metallic layer underneath and vacuum on top to calculate the average electrostatic potential, $V(x,y)$, generated by randomly distributed charges located within the four sulfur atomic layers of the tb-MoS$_2$ stack, which are schematically represented by red dots in Figure~\ref{fig:fig3}a. At distances $r \gg 2d$ (the thickness of tb-MoS$_{2}$) the radial dependence of the potential is $V(r)\sim1/r^{3}$, as expected for the dipole of the vacancy charge and its image in the graphite. See Methods section for more details.

To quantitatively compare the model with the experimental data, we assume that the band onsets in Figures~\ref{fig:fig2}b and c have a component proportional to the electrostatic disorder potential, $V(x,y)$, in addition to the dependence on the location within the moir\'e unit cell. We then compute the radially averaged autocorrelation functions, $\phi(r)$ which should reflect the radial dependence of the disorder potential $V(r)$. We fit $\phi(r)$ to $V(r)$ to determine the single fit parameter $n_{V}$. 

The solid curves in Figures~\ref{fig:fig3}b and c are the radial profiles of the onset map's autocorrelation functions associated with the valence and conduction bands, respectively. A positive autocorrelation is observed at small distances, consistent with long-ranged disorder. However a significant spatial modulation in both datasets is evident, which repeats approximately every moir\'e lattice constant and is due to the large variation of the onset energy within the moir\'e unit cell, which is not completely removed by radial averaging. To overcome this, we fitted the experimental onset energies in the regime $r>10$ nm to power laws (dashed lines), obtaining exponents $\alpha=-3.22$ for $\Gamma_{\text{v}}$ and $\alpha=-3.36$ for K$_{c}$, respectively. The exponents are in excellent agreement with the expectation $\alpha=-3$ i.e. $V(r)\sim1/r^{3}$ at large $r$. Figures~\ref{fig:fig3}d and e show the experimental $\phi(r)$ at small $r$ along with the model, and the power-law fit to the experimental $\phi(r)$ (dashed lines) for the the $\Gamma_{\text{v}}$ and K$_{\text{c}}$ onsets respectively. The experimental $\phi(r)$ is systematically lower at small $r$ than predicted from the long-range tail. This may be due to the fact that the onset of LDOS as measured by STS probes a property of the moir\'e superlattice, and may not measure disorder on length scales smaller than $\lambda_{M}\sim$ 20 nm.

The prefactor of the power-law fit allows us to determine $n_{\text{V}}$. We obtain two estimates, $n_{\text{V}} = 1 \times 10^{12} \text{cm}^{-2}$ and $n_{\text{V}} = 3 \times 10^{11} \text{cm}^{-2}$ (red solid curves) for the $\Gamma_{\text{v}}$ and K$_{\text{c}}$ onsets respectively. Remarkably, the fitted densities for $n_{\text{V}}$ obtained through our model lie within the reported V$_{S}$-associated defect-density range for mechanically exfoliated MoS$_2$ monolayers of $1\times 10^{11}\text{cm}^{-2}-1\times 10^{12}  \text{cm}^{-2}$~\cite{mos2_intr_dis_stm, lu2014bandgap, mcdonnell2014defect,exp_vac} (and below the typical reported defect-densities for CVD-grown crystals of $1\times 10^{12}\text{cm}^{-2}-1\times 10^{13}\text{cm}^{-2}$~\cite{defects_cvd_1,defects_cvd_2,defects_cvd_3}).

Finally, we model the effect of the electric field due to the areal density of charged defects $n_{\text{V}}$ on the moir\'e bands. We use the continuum model for tb-MoS$_2$ with $\theta = 0.95^\circ$ under the presence of an interlayer bias potential $V_z = 20$~meV, which corresponds to the electric potential associated with the higher estimated defect density $n_{\text{V}} = 1 \times 10^{12} \text{cm}^{-2}$. Details of the calculation are given in the Methods section. 

\begin{figure*}[ht!]
    \centering
    \includegraphics[width=1.0\textwidth]{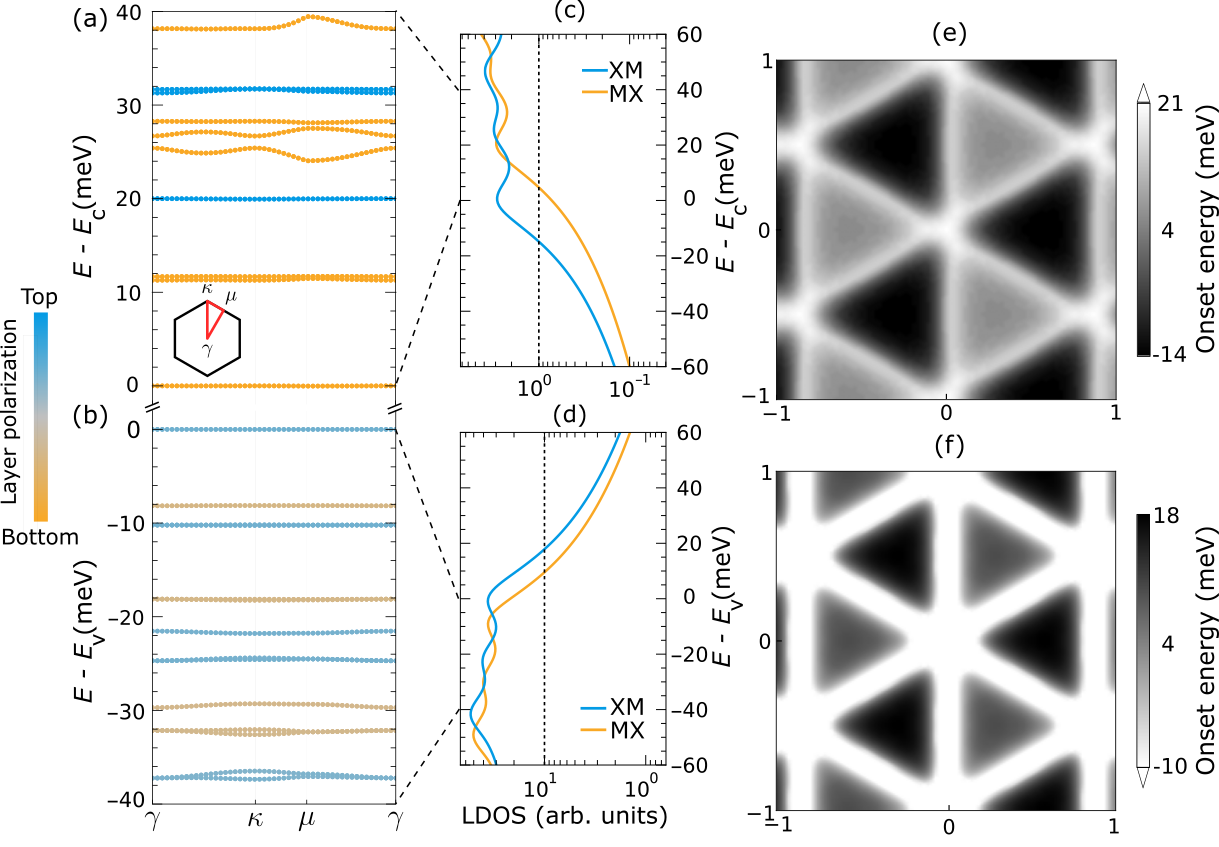}
    \caption{(a,b) Moir\'e conduction (a) and valence (b) minibands for tb-MoS$_{2}$ with $\theta = 0.95^\circ$ and a finite interlayer electric potential difference $V_z = 20$~meV along the mBZ path shown in the inset. The color indicates the layer polarization. (c,d) LDOS of conduction (c) and valence (d) bands, obtained assuming a Lorentzian broadening $\gamma = 10$~meV to account for charged disorder and thermal effects. The horizontal axis is in logarithmic scale. (e,f) Real-space maps (coordinates in units of the moir\'e length) of the onset energies at which the LDOS first crosses the dashed lines in (c,d) in the conduction (e) and valence (f) bands. Similar to Figures~\ref{fig:fig1}e, f, for some locations along the AA and DW regions, the color bars are not bounded in one direction (indicated by the arrow).}
    \label{fig:fig4}
\end{figure*}

Figure~\ref{fig:fig4} shows the results of our calculations. Figures~\ref{fig:fig4}a and b show the low-energy moir\'e bands for finite interlayer bias for the conduction (Figure~\ref{fig:fig4}a) and valence (Figure~\ref{fig:fig4}b) bands with their layer polarization indicated by the color scale. It can be seen that both valence and conduction bands become layer-polarized. Figures~\ref{fig:fig4}c and d show the average LDOS at the MX and XM stacking centers in the moir\'e cell, respectively, assuming a Lorentzian broadening $\gamma = 10$~meV to account for charged disorder and thermal effects. An onset shift of few tens of meV is observed between the spectra, similar to the experiment (Figures~\ref{fig:fig2}d and e). Figures~\ref{fig:fig4}e and f we show the calculated onset maps, defined as the energy where the LDOS reaches the value denoted by the dashed lines in Figures~\ref{fig:fig4}c and d. A visible color contrast between the triangular MX and XM regions within the moir\'e unit cell is seen, on the same energy scale as the experimental onset maps (Figures~\ref{fig:fig2}b and c).

\section{Discussion}

The results demonstrate that an out-of-plane electric potential consistent with the observed charge disorder can explain the unexpected LDOS onset shift between XM and MX domains of the same order of magnitude as that observed experimentally.

While the shift in LDOS onset with electric field is similar for valence and conduction bands, the mechanisms are somewhat different. For the lower conduction bands (K$_{c}$), the corresponding Wannier states are already layer-polarized even in the absence of an external field, as explained above. When an electric potential difference exists between the layers, the bands shift vertically in energy according to their dipole orientation; XM (MX) states accumulate charge on the top (bottom) layer. See SI Figure~S10b. The energy difference between the oppositely polarized conduction bands corresponds to the onset shift observed in the LDOS.

For the highest valence bands ($\Gamma_{\text{v}}$), the bands are equally distributed in both top and bottom layers (unpolarized) at zero interlayer bias (Figure~\ref{fig:fig1}d), and form a graphene-like bonding and anti-bonding manifold with layer pseudospin in the x,y plane. In an electric field, the bands acquire a layer polarization, developing a subband splitting whose magnitude is proportional to the electric-potential strength. This causes the MX states to accumulate electric charge on the top layer of the stacking, while the XM states do so on the bottom layer. See SI Figure~S10a. The polarization and shift in LDOS onset is reversed with respect to the conduction band, as observed experimentally (Figures~\ref{fig:fig2}d and e).

The calculated LDOS curves (Figures~\ref{fig:fig4}b, c) show no low-dispersion / flat band-related features, consistent with the averaged measurements (Figures~\ref{fig:fig2}d, e). This can be attributed to the intrinsic disorder observed in our experiments.

It is probable that other sources of electric field may also significantly perturb the electronic structure of tTMDs. Particularly in STM experiments the nearby metal tip, with different work function than the sample, may produce a built-in electric field which varies with bias voltage and which may locally greatly exceed the average electric fields considered in this work. This tip electric field is highly localized, potentially to a region smaller than one moir\'e unit cell, and further work is needed to understand the effect of such a highly localized field on the moir\'e bandstructure.

\section{Conclusions}

In conclusion, our results indicate that even modest charge disorder in monolayer MoS$_2$ (density $\sim$ 10$^{11}$ cm$^{-2}$, likely present in most exfoliated TMDs) can radically alter the character of the moir\'e bands in tTMDs at tens-of-meV energy scale. For example, the topmost moir\'e valence bands are transformed from a graphene-like manifold to two widely separated (much greater than bandwidth) completely layer-polarized bands. This poses profound implications for realizing certain flat band topologies in tTMDs, and indicates that care must be taken in interpreting experimental results.

\section{Methods}

\subsection{Experimental methods}

\subsubsection*{Sample fabrication}

STM measurements are performed on a tb-MoS$_2$ transferred onto a $\sim 30$~nm thick graphite flake for electrical contact. Sample was built using Si wafers coated with $285$~nm SiO$_2$. MoS$_2$ monolayers were obtained by mechanical exfoliation from high quality bulk crystals and subsequently stacked via a thin film of polycarbonate on a 6$\%$ chloroform solution using the tear-and-stack technique adapted from Ref.~\cite{cao2016superlattice}, aiming for a perfectly parallel stacking in order to obtain a near-zero twist angle. See SI Figure~S1 for further sample details.
The sample was then annealed in ultra high vacuum (UHV) conditions at $200^\circ$C for 2 hours and transferred to the STM chamber.

\subsubsection*{STM measurements}

STM measurements were performed using an electrochemically etched tungsten tip that was calibrated on clean Au(111) to the Shockley surface state.

Coarse positioning of the STM tip over the moir\'e region of interest was accomplished using a capacitance-based technique adapted from Ref.~\cite{navigation}.

STS grid measurements parameters for valence (conduction) bands are: $I_{\text{set}}=250$~pA, $45\times45$~nm$^2$ ($55\times55$~nm$^2$) with $40\times40$ pixels, $3.8$~mV ($5.1$~mV) bias step-size, 40~ms integration time, $20$~mV lock-in amplitude and frequency of 793~Hz.

All the measurements here reported were performed in UHV conditions at a temperature of $4.7$~K.

\subsection{Onset bias calculation}

The onset maps in Figures\ref{fig:fig2}b and c were calculated by finding the intersection between linear fittings for the band edges of log(dI/dV) curves and arbitrary noise-floor lines at $2.5\times10^{-2}$ (arb.units). See SI Figure~S2. 
Top region of these maps appears brighter as the intersection with that specific noise floor is not bounded for the energy scale at which the color contrast is more visible. 

The onset values presented in Figures~\ref{fig:fig2}d and e were found using the same method described above with arbitrary noise-floor lines at $5.25\times10^{-2}$ (arb.units) for the valence band (d) and at $1\times10^{-2}$ (arb.units) for the conduction band (e).
The uncertainty of the average onset bias was computed assuming statistically independent measurements as
$\sigma_{\mathrm{mean}}^2 = (\sum_i n_i)^{-2} \sum_i n_i \sigma_i^2$,
where $n_i$ and $\sigma_i^2$ denote the number of points and the variance associated with region $i$, respectively.

\subsection{Disorder density modeling}

The autocorrelation function of onset maps was obtained using \textit{numpy} library in Python $3.12$. 

The tb-MoS$_2$ is treated as two dielectric slabs, each of thickness $d\sim\SI{6}{\angstrom}$~\cite{al2016chemical} and monolayer MoS$_2$ out-of-plane relative electrical permittivity $\varepsilon_\text{r} = 5.5$~\cite{laturia2017dielectric} was used to calculate the electric potential across the stack.
The contribution of the dielectric slabs to the electric potential arises from an infinite ladder of image charges with alternating sign and decreasing magnitudes. The resulting image-charge amplitudes generates a rapidly converging geometric series that we truncate at $N=50$ terms. 

The presence of a graphite substrate in electrical contact with the MoS$_2$ bottom layer breaks the vertical inversion symmetry of the ideal free-standing model~\cite{zhang2014direct}. However, this is not considered in our model as the above-mentioned graphite Thomas-Fermi screening is orders of magnitude smaller than the moir\'e lattice constant in our sample.

\subsection{Theoretical Calculations}

\subsubsection{First-principles calculations}

First-principles density functional theory (DFT) calculations were performed to simulate bilayer MoS$_2$, in the rhombohedral (aligned) stacking, using the {\sc abinit} \cite{gonze2020,verstraete2025abinit} code, using the projecter augmented wave (PAW) method \cite{torrent2008implementation}.
{\sc abinit} employs a plane-wave basis set, which was determined using a kinetic energy cutoff of $1000$ eV. 
A Monkhorst-Pack $k$-point grid \cite{mp} of $12 \times 12 \times 1$ was used to sample the Brillouin zone. 
The PBE exchange-correlation functional was used \cite{pbe}, and the vdw-DFT-D3(BJ) \cite{becke2006simple} correction was used to treat the vdW interactions between the layers.

To sample the different possible local environments, we used $37$ stacking configurations that are not related by symmetry on a triangular grid (see Figure\ S4), which is compatible with the $C_{6v}$ symmetry of the adhesion potential. 
For each relative stacking, a geometry relaxation was performed to obtain the equilibrium layer separation, while keeping the in-plane atomic positions fixed, using a force tolerance of 1 meV/Å. 
Band structure calculations were then performed for each stacking, using the relaxed geometries.

\subsubsection{Continuum model of tMoS$_2$}

Here we briefly describe the continuum model used to calculate the band structure of tMoS$_2$. A detailed derivation is provided in SI.

We construct an effective Hamiltonian for aligned bilayer 2H MoS$_2$ near $\Gamma_v$ (top of the valence band) and $K_c/K_c'$ (bottom of the conduction band) as a function of the layer stacking $\bm \phi$. 
In a layer basis, this Hamiltonian can be approximated as
\begin{equation} \label{eq:Hkp}
    H_Q(\bm k,\bm \phi) = \begin{bmatrix}
        \tfrac{\hbar^2 k^2}{2m_Q} + \epsilon_Q(\bm \phi) & t_Q(-\bm \phi) \\ t_Q(\bm \phi) & \tfrac{\hbar^2 k^2}{2m_Q} + \epsilon_Q(-\bm \phi)
    \end{bmatrix},
\end{equation}
where $\bm k = (k_x, k_y)$ is the Bloch momentum and $m_Q$ is the effective mass at valley $Q=\Gamma_v,\text{K}_c$, $\epsilon(\bm \phi)$ is the intralayer potential and $t(\bm \phi)$ is the interlayer tunneling amplitude.
Eq.~\eqref{eq:Hkp} is parametrized using DFT calculations (see SI).

For small twist angles, the moir\'e pattern varies slowly on the interatomic scale, and we use the local stacking approximation~\cite{jung_origin_2015}:
\begin{equation} \label{eq:lsa}
    \begin{aligned}
    \bm \phi(\bm r) & = R(\theta/2) \bm r - R(-\theta/2) \bm r + \bm u(\bm r) \\
    & = \frac{a}{L} \hat z \times \bm r + \bm u(\bm r),
    \end{aligned}
\end{equation}
where $R(\theta)$ is the rotation matrix for an angle $\theta$ counterclockwise about the $z$ axis, $L = a/2\sin(\theta/2)$ is the moir\'e lattice constant, and $\bm u = \bm u_1 - \bm u_2$ is the acoustic displacement field arising from lattice reconstruction.

For example, at $\Gamma_v$ the continuum Hamiltonian can be written as $H_{\Gamma} = \sum_{s = \uparrow,\downarrow} \int d^2 \bm r \, \psi_s^\dag(\bm r) \mathcal H_{\Gamma} \psi_s(\bm r)$ with
\begin{equation} \label{eq:continuum}
    \mathcal H_\Gamma =
    -\tfrac{\hbar^2 \nabla^2}{2m_\Gamma} \sigma_0 + \begin{bmatrix}
        \epsilon(\bm r) + \tfrac{V_z}{2} & t(\bm r) \\ t(\bm r) & \epsilon(-\bm r) - \tfrac{V_z}{2}
    \end{bmatrix},
\end{equation}
where $\psi_s = ( \psi_{s1}, \psi_{s2} )^\top$ are fermion field operators in layer space that obey the usual anticommutation relations. 
We also include an interlayer bias $V_z$ which corresponds to an electric displacement.

The moir\'e bands are obtained by writing down a Bloch \textit{ansatz} for the moir\'e continuum theory in terms of a plane-wave expansion.

\subsubsection{Local density of states}

To compare our theory to the STM measurements described in the main text, we compute the local density of states (LDOS)
\begin{equation}
    \rho(\bm r, E) = \sum_{n,\bm k} \delta(E - E_{n\bm k}) |\psi_{n,\bm k}(\bm r)|^2,
\end{equation}
where $n$ is a band index (including spin, as well as valley for $K_c/K_c'$). Here the total wave function squared includes a sum over both layers and reciprocal moir\'e vectors of the plane-wave expansion.
To account for disorder, we introduce a phenomenological broadening parameter $\gamma$ and replace the delta function with a Lorentzian:
\begin{equation}
    \delta(E - E_{n\bm k}) \rightarrow \frac{1}{\pi} \frac{\gamma}{(E - E_{n\bm k})^2 + \gamma^2},
\end{equation}
with a full width at half maximum given by $2\gamma$.

\subsubsection{Lattice Reconstruction}

In moir\'e materials, reconstruction is driven by a competition between intralayer elastic energy, which prefers the rigid configuration, and interlayer stacking energy, which favors expanding regions of favorable stacking configurations.
We model this atomic reconstruction of the moir\'e pattern with continuum elasticity \cite{nam_lattice_2017,carr_relaxation_2018,bennett_theory_2022,debeule_theory_2025}, which is a good approximation when the moir\'e period $L \gg a$ and the local stacking varies slowly compared to the interatomic scale.
The continuum model, described in detail in SI, is described by the free energy
\begin{equation}\label{eq:F-relax}
    \mathcal F = \mathcal F_\text{elas}[\bm u_1] + \mathcal F_\text{elas}[\bm u_2] + \mathcal F_\text{stack}[\bm \phi] 
\end{equation}
where $\mathcal F_\text{elas}$ is the elastic energy, $\bm u_{1,2}(\bm r)$ are the displacements of layers 1 and 2, $\mathcal F_\text{stack}$ is the stacking/adhesion energy, and $\phi(\bm r) = \theta \hat z \times \bm r + \bm u_1(\bm r) - \bm u_2(\bm r)$ is the relative stacking between the layers, in the local stacking approximation. Here, Eq.~\eqref{eq:F-relax} was parametrized using first-principles calculations as described in SI. The effects of lattice relaxation on the band structure are accounted for in the continuum model.

\section*{Supporting information}

The Supporting Information is available free of charge at: Supporting\_Information.pdf. It includes:

\begin{enumerate}
    \item Details on the experimental sample and determination of the onset bias, including photographs, and
    \item theory calculations, including structural relaxation and the relaxed electronic continuum model used to calculate the local density of states as a function of interlayer bias for both the $\Gamma_v$ and $K_c / K_c'$ moir\'e bands.
\end{enumerate}

\section*{Acknowledgment}

PRV, THYV, KX, SA, MTE and MSF acknowledge support from ARC Discovery Project DP200101345. DM, SA, MTE and MSF acknowledge funding support from the ARC Centre for Future Low Energy Electronics Technologies (FLEET) CE170100039. 
This work was performed in part at the Melbourne Centre for Nanofabrication (MCN) in the Victorian Node of the Australian National Fabrication Facility (ANFF).
D.B.~acknowledges support from the NTU Startup Grant (Award Number 025661-00003).
H.G-H. acknowledges financial support from the Spanish State Research Agency under grant Ram\'on y
Cajal fellowship RYC2021-031050-I


\clearpage  
\printbibliography


\end{document}